# Parallelize Bubble and Merge Sort Algorithms Using Message Passing Interface (MPI)

Zaid Abdi Alkareem Alyasseri [1], Kadhim Al-Attar [2], Mazin Nasser [2] and ISMAIL [3]

[1]Faculty of engineering, university of Kufa, Iraq
[2]School of computer science, USM, Malaysia
Zaid.alyasseri@uokufa.edu.iq

***Abstract*:** Sorting has been a profound area for the algorithmic researchers and many resources are invested to suggest more works for sorting algorithms. For this purpose, many existing sorting algorithms were observed in terms of the efficiency of the algorithmic complexity. In this paper we implemented the bubble and merge sort algorithms using Message Passing Interface (MPI) approach. The proposed work tested on two standard datasets (text file) with different size. The main idea of the proposed algorithm is distributing the elements of the input datasets into many additional temporary sub-arrays according to a number of characters in each word. The sizes of each of these sub-arrays are decided depending on a number of elements with the same number of characters in the input array. We implemented MPI using Intel core i7-3610QM ,(8 CPUs),using two approaches (vectors of string and array 3D) . Finally, we get the data structure effects on the performance of the algorithm for that we choice the second approach.

## 1. Introduction

Sorting is one of the most common operations perform with a computer. Basically, it is a permutation function which operates on elements [1]. In computer science sorting algorithm is an algorithm that arranges the elements of a list in a certain order. Sorting algorithms are taught in some fields such as Computer Science and Mathematics. There are many sorting algorithms used in the field of computer science. They differ in their functionality, performance, applications, and resource usage[2].We are going to give a brief introduction of the most popular sorting algorithms.

## 2. Bubble Sort Algorithm

Bubble sort is the oldest, the simplest and the slowest sorting algorithm in use having a complexity level of O(n2). Bubble sort works by comparing each item in the list with the item next to it and swapping them if required. The algorithm repeats this process until to make passes all the way through the list without swapping any items. Such a situation means that all the items are in the correct order. By this way the larger values move to the end of the list while smaller values remain towards the beginning of the list. It is also used in order to sort the array such like the larger values comes before the smaller values [1]. In other words, all items are in the correct order. The algorithm's name, bubble sort, comes from a natural water phenomenon where the larger items sink to the end of the list whereas smaller values "bubble" up to the top of the data set [2]. Bubble sort is simple to program but it is worse than selection sort for a jumbled array. It will require many more component exchanges, and is just good for a pretty well ordered array. More importantly bubble sort is usually the easiest one to write correctly [4].

### 3.1 Bubble Sort as sequential

The sequential version of the bubble sort algorithm is considered to be the most inefficient sorting method in common usage. In the sequential code we implement two approaches; the difference between them is the data structure. 1) Implemented based on vectors of strings; 2) implemented based on array char 3D.
The performance of the sequential code for two datasets which has been tested for 10 times to get the average is shown in table 1.

Table 1: Shows the performance of sequential code

| Test/ Sec | Dataset1 | Dataset2 |
|---|---|---|
| Test1 | 6.695 | 188.185 |
| Test2 | 6.624 | 188.194 |
| Test3 | 6.615 | 188.247 |
| Test4 | 6.694 | 188.169 |
| Test5 | 6.531 | 188.348 |
| Test6 | 6.654 | 188.289 |
| Test7 | 6.654 | 188.154 |
| Test8 | 6.614 | 188.512 |
| Test9 | 6.646 | 188.343 |
| Test10 | 6.672 | 188.181 |
| Ave | 6.639 | 188.262 |

## 3.2 Bubble Sort as Parallel

One of the fundamental problems of computer science is ordering a list of items. There are a lot of solutions for this problem, known as sorting algorithms. Some sorting algorithms are simple and intuitive, such as the bubble sort, but others, like quick sort, are extremely complicated but produce lightning-fast results. The sequential version of the bubble sort algorithm is considered to be the most inefficient sorting method in common usage. In this assignment we want to prove that how the parallel bubble sort algorithm is used to sort the text file parallel and will show that it may or may not better than the sequential sorting algorithm. The old complexity of the bubble sort algorithm was O(n2), but now we are using the complexity for bubble sort algorithm n(n-1)/2. Algorithm 1 ( in chapter 1) shows the code for the bubble sort algorithm. As usually in parallelism we can decompose the data or functions or sometimes both, but after we are studying the bubble sort we reached to the point, that it is more suitable for data decomposition. So we suggest during the implementation of MPI is to assign each vector to individual process and during the implementation of MPI we divide each vector to different chunks and assign each chunk to individual process, as a result we merge all the executed chunks for the solution of the problem.

## 3. Methodology

There are two types of text file dataset have been provided in this paper (*HAMLET, PRINCE OF DENMARK by William Shakespeare*), which are different in size and length. The first dataset is equal *(190 KB)* and the second one is equal *(1.38 MB)* taken from (http://www.booksshouldbefree.com/).

**Pre-processing**

We sort the datasets using the bubble sort algorithm in three phases. In the first phase we are removing / ignoring the special characters from the text file. In the second phase we convert the text file to array of list (vectors of string) based on the length of characters, all shorter words come be for longer words. In the third phase we sort each vector of string by arranging in the alphabetic order using the bubble sort algorithm. The data set has been tested 10 times to get the average. The text files used for this assignment are in English. The sequential performance has been implemented on ASUS A55V, Windows 7 Home Premium 64-bit with Intel core i7-3610QM, (8 CPUs), 2.30GH and 8 GB of RAM.
Table 2: shows the time of pre-processing for Dataset 1 and 2

| Function | Dataset 1 (190KB) | Dataset2 (1.38MB) |
|---|---|---|
| **Preprocessing to remove the special characters from a text file** | 0.265 second | 1.154 second |
| **Preprocessing on file by using bubble sort based on alphabetic order** | 274.528 second | 274.528 second |
| **Preprocessing on file by using bubble sort based on the length of the word** | 230.215second | 230.215second |
| **Bubble sort on array of vector each vector has the same length of the word** | 42.589 second | 1620.44 second |

## 4. Message Passing Interface (MPI)

Message Passing Interface (MPI) is a specification for a standard library for message passing that was defined by the MPI Forum, a broadly based group of parallel computer vendors, library writers, and application specialists(William Gropp). MPI is a specification not an implementation. The main purposeofMPI is to develop a widely used standard for writing message passing programs and achieving practical, portable, efficient and flexible standard for message passing. MPI can be supported and implanted by C, C++, Fortran-77, and Fortran-95. MPI is a mixture of functions, subroutines, and methods and also have different syntax from other languages.MPI is a de-factor standard for message-passing software used for the development of high-performance portable parallel applications. The MPI specification has been implemented for a wide range of computer systems from clusters of networked workstationsrunning general purpose operating systems (UNIX, Widows NT) to high performance computer systems such as CRI T3E, CRI C90, SGI Power Challeng, Intel Paragon, IBM SPI, SP2 etc. [6].

In parallel programming the problem will be divided intosubproblems or processes and assign to different processors. The rule of MPI is to establish communication among the processes, hence MPI is used for distributed memory not for the shared memory. This means that the system worksonly forsending and receiving the data with their own memory.

Different versions of MPI are used such as: MPI 1.0, MPI 1.2, MPI 2.0, MPI 1.3, MPI 2.1, and MPI 2.2, these versions have different functionality and facilities.

### MPI Implementation

We are implementing MPI code using *Master/slave* communications where the first thing we have calculates the size of chunks equal the length of the array divided by the number of processors *"Whenever the chunk size small the parallel time is better"* . After that the master will send the data to the slaves and each slave will do a bubble sort on own data. At the end of this step the slave receives the result to master. The final step merges all the chunks and prints the results. Figure 1 described the proposed method for sorting Dataset 1 and 2.

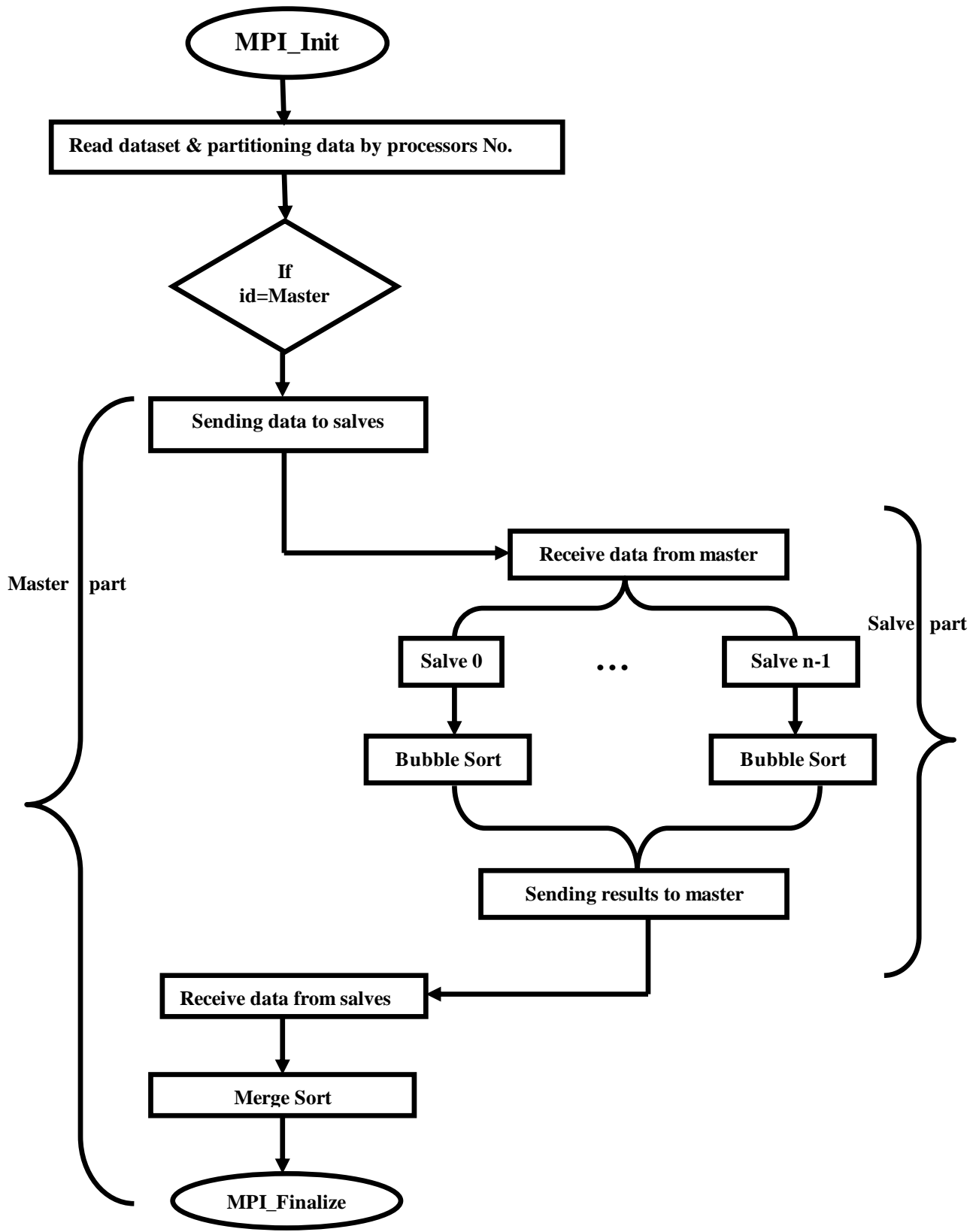

**Figure 1.** Shows the proposed method of MPI

## 5. Results and Analysis

In the MPI result, we obtained the speedup always is increasing when we increase the number of threads because the implemented MPI using (**bubble and merge sort** ).
Where the **chunk size = array size / no. of processors.**
So when we increase the no. of processors that will lead to reduce the chunk size, thus will reduce the parallel time. For example, array size= 40 and we have 4 processors so if we run the program use 2 the chunk size = 20 and when we used 4 processors. Now the chunk size became equal 10 and the program run with chunk size 10 is done faster than with 20 size.

| Thread No. | Data1 Time/ Sec | Data2 time/ Sec | Speed up Data 1 | Speed up Data 2 | Efficiency Data 1 | Efficiency Data 2 |
|---|---|---|---|---|---|---|
| 1 | 4.703 | 181.908 | 1 | 1 | 100 | 100 |
| 2 | 1.258 | 49.836 | 3.738 | 3.650 | 186.924 | 182.507 |
| 4 | 0.507 | 20.497 | 9.276 | 8.875 | 231.903 | 221.871 |
| 6 | 0.308 | 15.277 | 15.269 | 11.907 | 254.491 | 198.455 |
| 8 | 0.237 | 13.804 | 19.844 | 13.178 | 248.049 | 164.724 |
| 10 | 0.22 | 13.84 | 21.377 | 13.144 | 213.773 | 131.436 |
| 16 | 0.189 | 12.946 | 24.884 | 14.051 | 155.522 | 87.821 |

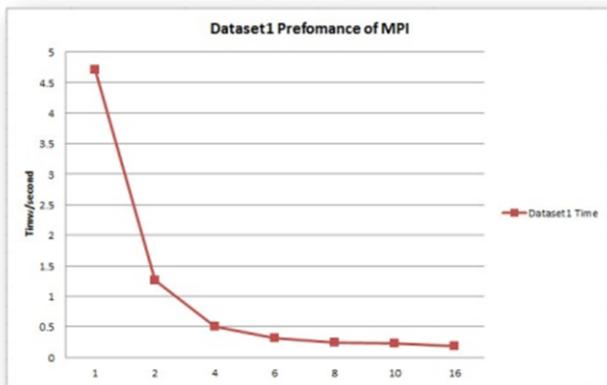

**Figure 2.** shows the performance of MPI with Dataset1

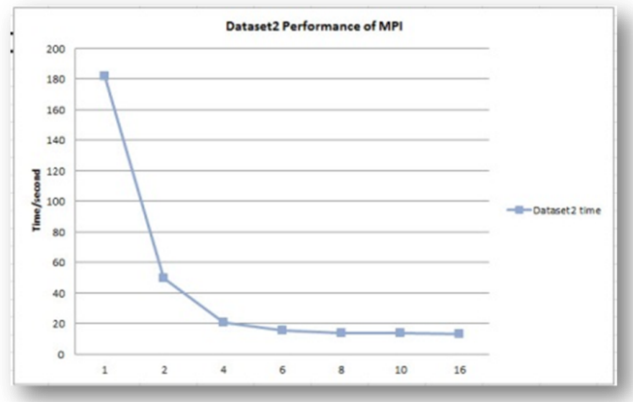

**Figure 3.** shows the performance of MPI with Dataset2

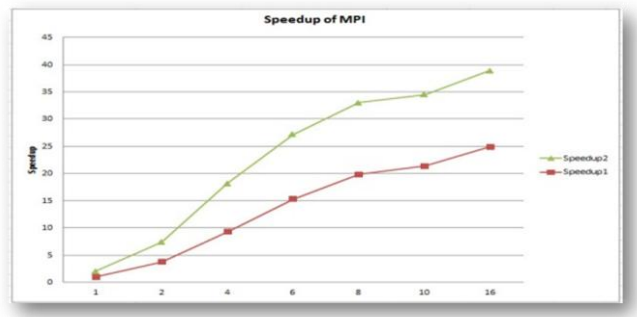

**Figure 4.** Shows the speedup of dataset1 and 2 using MPI

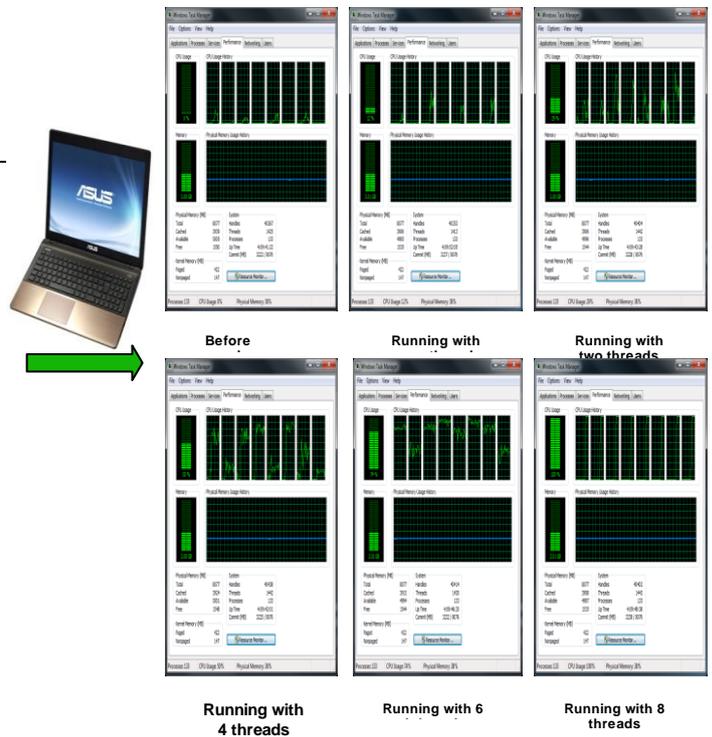

**Figure 5.** Shows running algorithm on 1 to 8 threads

## 6. Conclusions

In this paper we implemented the bubble sort algorithm using massage passing interface (MPI). The proposed work tested on two standard datasets (text file) with different size taken from (*HAMLET, PRINCE OF DENMARK by William Shakespeare*), (http://www.booksshouldbefree.com/). We implemented MPI using ASUS A55V, Operating System: Windows 7 Home Premium 64-bit Processor: Intel core i7-3610QM ,(8 CPUs), 2.30GH ,memory: RAM 8 GB. We were finding the data structure effects on the performance, where this is clear in sequential code 1 and 2. In OpenMP, increasing the number of threads more than an actual core number . In MPI, increasing the number of threads it will increase speedup, because the approach of MPI which implements here (bubble and merge sort).
.